\documentstyle{article}
\def\bichpr{\hoffset=-20truemm
\voffset=-30truemm
\textwidth=16 truecm
\textheight=24 truecm }

\bichpr
\begin{document}
\title{Towards a 3D reduction of the N-body Bethe-Salpeter equation.} 
\author{ J. Bijtebier\\
 Theoretische Natuurkunde, Vrije Universiteit Brussel\\ Email:
jbijtebi@vub.ac.be}
\date{}
\maketitle

\section{Introduction.} The Bethe-Salpeter equation is the usual tool
for computing bound states of relativistic particles. The principal difficulty of
this equation comes from the presence of N-1 (for N particles) unphysical degrees of
freedom: the relative time-energy degree of freedom. In the two-body problem, the
relative energy is usually eliminated by replacing the free two-body propagator by an
expression combining a delta fixing the relative energy and a 3D propagator. The
exact equivalence (in what concerns the physically measurable quantities of the
pure two-fermion problem) with the original Bethe-Salpeter equation can be
obtained by recuperating the difference with the original free propagator in a
series of correction terms to the 3D potential. It is not possible to generalize 
this constraining propagator-based reduction method to three or more particles,
because of the unconnectedness of the two-body terms of the Bethe-Salpeter kernel,
which are in fact the more important terms and often the only ones to be considered.
\par A less often used 3D reduction method is based on the replacement of the
Bethe-Salpeter kernel by an "instantaneous" (i.e. independent  of the relative
energy) approximation (kernel-based reduction). In this case, the resulting 3D
potential is not manifestly symmetric (i.e. hermitian when the total energy on which
it depends is treated as a parameter).  In the two-fermion problem, we obtained a
symmetric 3D potential by performing a supplementary series expansion at the 3D
level and combining it with the first 3D reducing expansion. We found that the
starting instantaneous approximation of the Bethe-Salpeter kernel disappears from
the final 3D potential.  In fact, this potential can be obtained  directly by a new
integrating propagator-based reduction method, in which the relative energy is
integrated on, instead of being fixed by a $\,\delta-$fonction (or constraint).\par 
This integrating propagator-based reduction can easily be generalized to a
system of N particles, consisting in any mixing of bosons and fermions
\cite{1,2}.

\section{Inhomogeneous and homogeneous Bethe-Salpeter equations for 2 fermions:}
$$G=G^0+G^0KG,\qquad\Phi = G^0 K \Phi$$
\indent $\Phi:$ \qquad\qquad\qquad\qquad\, Bethe-Salpeter  amplitude\par
$K:$ \qquad\qquad\qquad\qquad\, Bethe-Salpeter kernel\par 
\qquad\qquad\qquad \qquad\qquad\qquad(must give $\,G\,$ via
the inhomogeneous equation)
\par
$G\equiv G^0+G^0TG^0:$\qquad Full propagator (Feynman graphs) \par
$G^0:$\,\,\,\qquad \qquad \qquad \qquad Free propagator:  
$$G^0=G^0_1G^0_2,\qquad G^0_i ={1\over \gamma_i\cdot
p_i\,-\,m_i\,+i\epsilon}= {1 \over p_{i0}-h_i+i\epsilon h_i}\,\beta_i$$
$$h_i = \vec \alpha_i\, . \vec p_i + \beta_i\, m_i\qquad (i=1,2)$$
The self-energy parts of the propagator were transferred to the kernel. We shall
neglect them here for simplicity. $\,K$ is then the sum of the irreducible
two-fermion Feynman graphs.\hfill\break\par\noindent
\vfill\break
Notations for the following:
$$ P = p_1 + p_2\ , \qquad p = {1 \over 2} (p_1 - p_2)$$
$$ E = E_1+E_2,\quad E_i=\sqrt{h_i^2}=(\vec
p_i^2+m_i^2)^{1\over 2}.  $$
$$\Lambda^+=\Lambda_1^+\Lambda_2^+,
\qquad \Lambda_i^+={E_i+ h_i\over 2E_i},\qquad\beta=\beta_1\beta_2$$

\section{3D reduction by expansion around a positive-energy 
instantaneous approximation of K.}

Write\quad $\,K\,=\,K^0\,+\,K^R\,$ with
$\,K^0\!=\!\Lambda^{+}\beta K^0\Lambda^{+}\,$ (positive-energy) and
$\,K^0(p'_0,p_0)\,$ independent of $\,p'_0,p_0\,$ (instantaneous). The
Bethe-Salpeter equation becomes
$$\Phi\,=\,G^0K^0\Phi\,+\,G^0K^R\Phi\qquad\to$$
$$\Phi\,=\,(1-G^0K^R\,)^{-1}G^0K^0\Phi\qquad\to\qquad\Phi\,=\,(G^0+G^{KR})K^0\Phi$$ 
with
$$G^{KR}\,=\,G^0K^R(1-G^0K^R)^{-1}G^0.$$
Integrate with respect to $\,p'_0\,$ and apply $\,\Lambda^+\quad\rightarrow$
\quad 3D equation: 
$$\psi\,=\,(\,g^0+g^{KR}\,)\,V^0\,\psi$$
with
$$\Lambda^+\int dp_0 G^0(p_0)\,=\,-2i \pi\Lambda^+\, g^0\,  
\beta,\qquad g^0\,=\,{1 \over P_0-E+i\epsilon}$$
$$\psi\,=\,\Lambda^+\int dp_0\,\Phi(p_0)\,,\qquad
V^0\,=\,-2i\pi\,\beta\,K^0\,,$$
$$g^{KR}\,=\,{-1\over2i\pi}\,\Lambda^+\int
dp'_0dp_0\,G^{KR}(p'_0,p_0)\,\beta\label{44}\,\Lambda^+.$$

\section{Render the potential symmetric.}
The 3D potential $\,(g^0)^{-1}(\,g^0+g^{KR}\,)\,V^0\,$ is not
symmetric. In Phillips and Wallace's method \cite{3}, one computes $\,K^0\,$ in
order to make $\,g^{KR}\,$ vanish. Here, we shall write
$$g^{KR}=g^0\,T^{KR}g^0$$
$$\to\qquad\psi=(1+g^0T^{KR})\,g^0V^0\psi\qquad \to\qquad
(1+g^0T^{KR})^{-1}\psi\,=\,g^0V^0\psi$$
$$\to\qquad\psi\,=\,\big[\,g^0V^0+1-(1+g^0T^{KR})^{-1}\,\big]\,\psi
\qquad\to\qquad\psi=\,g^0\,V\,\psi$$
with $$\,V=V^0+T^{KR}(1+g^0T^{KR})^{-1}.$$
This potential $\,V\,$ is now symmetric. \vfill\break
\section{Expand $T^{KR}$ and recombine the series. }
$$T^{KR}\,=\,<K^R(1-G^0K^R)^{-1}>$$ with
$$<A>={1\over-2i\pi}\,\Lambda^+(g^0)^{-1}\int
dp'_0dp_0\,G^0(p'_0)A(p'_0,p_0)G^0(p_0)\,\beta\,\Lambda^+(g^0)^{-1}.$$
This leads to
$$V\,=\,<K^0>+<K^R(1-G^0K^R)^{-1}>(1+g^0<K^R(1-G^0K^R)^{-1}>)^{-1}$$
$$\,=\,<K^0+K^R(1-G^0K^R)^{-1}(1+>\!g^0\!\!<K^R(1-G^0K^R)^{-1})^{-1}>$$
$$\,=\,<K^0+K^R(1-G^0K^R+>\!g^0\!\!<K^R)^{-1}>\,=\,
<K^0+K^R(1-G^RK^R)^{-1}>$$
with the definitions
$$G^R\,=\,G^0-G^I,\qquad G^I=\,\,\,>g^0\!\!<.$$
Less formally:
$$G^0(p'_0,p_0)\,=\,G^0(p_0)\,\delta(p'_0-p_0),\qquad
G^I(p'_0,p_0)\,=\,G^0(p'_0)\,\beta\,{\Lambda^+
\over-2i\pi\,g^0}\,G^0(p_0).$$
but $$\,\,K^0G^R=G^RK^0=0\,\,\to\,\,K^R(1-G^RK^R)^{-1}=-K^0+K(1-G^RK)^{-1}
\,\,\to$$
$$V\,=\,<\,K\,(1-G^RK\,)^{-1}\,>\,=\,<K>+<KG^RK>+\cdots$$
$$=\,<K>+\,\{<KG^0K>-<K>g^0<K>\,\}+\cdots$$
In the relative-energy integrals, $\,-G^I\,$ cancels the leading term coming
from $\,G^0.$ \par\noindent
Good surprise: $\,V\,$ does not depend on the initial choice of $\,K^0\,$
anymore. 

\section{We made in fact an integrating propagator-based reduction.}
Our final 3D equation could also be obtained directly from the Bethe-Salpeter
equation by performing an expansion around an approximation
$\,G^I\,$ of the propagator $\,G^0\,$ (\,$\to\, $integrating propagator-based
reduction instead of the constraining propagator-based reduction using
$\,\delta-$functions).

\section{We could start with the equal-times retarded propagator.}
 Following \cite{4} (\cite{5} in the three-body case), we could also start by
taking the retarded part of the full propagator at equal times. In momentum
space, it is
$$g\,=\,g^0\,+\,g^0<T>g^0.$$
The corresponding 3D potential is
$$V\,=\,<T>(1+g^0<T>)^{-1}.$$
Writing then the expansion $\,T=K(1-G^0K)^{-1}\,$ and recombining the series
gives $\,V\!=\,<K\,(1-G^RK\,)^{-1}>\,$ again.  
 Note that
$\,T\,$ and $<T>$ are both proportional to the physical scattering
amplitude when the initial and final fermions are on their positive-energy mass
shell.
\vfill\break 

\section{Generalization to systems of N particles.}
 Our 3D reduction method (as established in section 5 or section 6's way)
can be easily generalized to systems consisting in any number of fermions
and/or bosons.\par Here we shall consider only the case of N fermions. The
writing of the Bethe-Salpeter equation and of the final 3D equation remains the
same:
$$\Phi\,=\,G^0K\,\Phi\qquad\to\qquad\psi\,=\,g^0\,V\,\psi$$
$$V\,=\,<K(1-G^RK)^{-1}>,\qquad G^R\,=\,G^0\,-\,\,\,\,>\!g^0\!\!<,\qquad 
g^0\,=\,{1\over P_0-E+i\epsilon} $$
with a trivial generalization of some notations:
$$\Lambda^+=\Lambda^+_1\cdots\Lambda^+_N,\qquad \beta=\beta_1\cdots\beta_N,$$$$
P_0=p_{01}+\cdots+p_{0N}\qquad E=E_1+\cdots +E_N$$
$$<A>={1\over(-2i\pi)^{N-1}}\,\Lambda^+(g^0)^{-1}\int
dp'_0dp_0\,G^0(p'_0)A(p'_0,p_0)G^0(p_0)\,\beta\,\Lambda^+(g^0)^{-1}$$
$$dp_0\,=\,dp_{01}\cdots dp_{0N}\,\delta\,(\,p_{01}+\cdots +p_{0N}\,-\,P_0\,).$$
\hfill\break\par\noindent
Expressions of $\,K\,$ for $\,N\ge3:$ \hfill\break\par\noindent
$N=3:$
$$ K=K_{12}(G_{03})^{-1}+K_{23}(G_{01})^{-1}+K_{31}(G_{02})^{-1}
+K_{123}.$$ 
$N=4:$
$$ K\,=\,K_{12,34}\,+\,K_{13,24}\,+\,K_{14,23}$$
$$+\,\,\,K_{123}\,(G^0_4\,)^{-1}\,+\,K_{124}\,(G^0_3\,)^{-1}\,+\,K_{134}\,
(G^0_2\,)^{-1}\,+\,K_{234}\,(G^0_1\,)^{-1}\,$$ 
$$+\,\,\,K_{1234}\,,$$ 
with
$$K_{12,34}\,=\,K_{12}\,(G^0_3\,G^0_4\,)^{-1}
\,+\,K_{34}\,(G^0_1\,G^0_2\,)^{-1}\,-\,K_{12}\,K_{34},\qquad etc...$$
The counter-term  $\,K_{12}\,K_{34}\,$ cancels the double-countings
which would come from the fact that two graphs containing
respectively $\,K_{12}\,K_{34}\,$ and
$\,K_{34}\,K_{12}\,$ in the expansion of $\,G\,$ must be taken only once
\cite{6,7,8,2}.
\hfill\break\par\noindent
$N\ge5:$\qquad Very complicated. For $\,N\ge5\,$ (perhaps  even for $\,N=4\,$) we
suggest to bypass the Bethe-Salpeter equation by writing
$V\!=\,<T>(1+g^0<T>)^{-1}\,$ without expanding $\,T\,$ in terms of $\,K,\,$ and
sorting the contributing graphs by increasing number of vertexes.

\end{document}